# Reuse of existing applications during the development of Enterprise Portals integrating Web Services

M. Gharzouli
*Lire laboratory, Department of Computer Science, Mentouri University of Constantine, 25000, Algeria*
gharzouli_mohamed@yahoo.fr

M. Boufaida,
*Lire laboratory, Department of Computer Science, Mentouri University of Constantine, 25000, Algeria*
boufaida@hotmail.com

L. Seinturier
*LIFL UMR CNRS 8022 Bâtiment INRIA Haute-Borne 59655 Villeneuve d'Ascq Cedex France*
Lionel.Seinturier@lifl.fr

**Abstract**

*During these last years, the use of the web technologies in the enterprises becomes an essential factor to define a new business model. Among these technologies, web services and enterprise portals have gathered to integrate existing heterogeneous systems such as e-commerce, e-services hub and e-learning. However, the design and the modeling of the portals, on the one hand, and their integration with the existing applications, on the other hand, are still two points open for discussion. The first problem is related to the development of the lifecycle that can be used for designing and modeling the enterprise portals. For the second problem, which is the integration of the existing applications, the discussion is intended towards the use of technologies based on web services. In This paper, we present a software engineering solution for the development of Web services-based enterprise portals.*

**Keywords:** Development process, Enterprise Portals, Reuse of existing application, Web services.

## 1. Introduction

The concept of Enterprise Portal (EP) has received much attention since the success of public portals on the Web like Yahoo! and MSN. It is increasingly being used to refer to vertical web sites that feature personalization/customization, cross-platform usability, distributed access, management, and security of information and services within a particular enterprise/industry, thus the so called enterprise, corporate or vertical portals [4]. Portal technology can be targeted towards external users (B2C and B2B) or/and internal users (B2E). It is a tool that incorporates on only one personalized screen, according to the profile of each user, all information and applications he needs to work with [1]. The EP decreases the spent time of searching information with giving the possibility to reach separate sources of information [12].

Progressively, portals have become a need for many famous companies like IBM, Microsoft and Plumtree. These later offer packaged solutions on a variety of platforms. However, these packages lack interoperability in general and synchronization among the various components in particular. Furthermore, they are very expensive [17]. For this reason, the common enterprises choose the "home development" of the portal solution. Moreover, one of the key properties of the portal is that it is built of heterogeneous applications. So, it is necessary to use a tool that permits the interoperability between the portal and the different applications.

In this direction, the use of Web services is considered to be a good solution for creating a homogeneous environment [11]. The web services architecture is based on three elements: the service provider, the discovering agency and the service requestor [13], [10], [5]. If a company wants to develop a portal based on web services architecture, it defines the role of the service provider and those of the service requestor. So, the development of such a portal requires firstly the preparation and the construction of its contents by re-using existing applications.

In this paper, we describe a methodology for the development of the web services for the enterprise portals with re-using existing applications. This methodology explains how we can develop initially the web services and then how to consume them through the portal.

In the following, we present an important aspect that is related to the use of interoperable web services for computational portals. After having presented the various steps of our methodology, we provide a comparison between some related works and our approach and we finish with a conclusion and some prospects.

## 2. Web services for Enterprise Portals

Recently, Web services and enterprise portals have gathered to integrate heterogeneous systems such as *e-*commerce, e-services hub and e-learning [16]. To ensure the interoperability in the various types of portals, Web services form an adequate and effective solution. The portal needs to reach heterogeneous information sources and re-uses existing applications. Web services answer these constraints because they offer a high degree of interoperability. The three basic standards WSDL, UDDI and SOAP make possible the description, the publication and the invocation of Web services [2], [6], [7], [10].

The basic interaction of the web services for a portal is based on a separation between the server that manages the user interface and the one that uses a particular service [11]. In the case of an enterprise portal, the SOAP server can be internal i.e. located on an Intranet. In this case, the provided services can be invoked via SOAP by using the local network area. If the SOAP server is external, the interaction is made via an Extranet or the Internet network.

## 3. An integration process of web services in the Enterprise Portals

Currently, there are several products that are offered by various editors of EP solutions. However, each editor has his own development methodology. For the internal development, the enterprises use generally the methods and the processes for the modeling of the web applications. But the solution, which is the most common, is to use RUP for the process and UML for the notation [8], [9].

The difficulty that arises for these methods is that they are oriented to the development of new applications. However, the enterprise portals reuse generally existing applications.

In the following, we give the great steps to be followed for the development of an enterprise portal providing a single and personalized access point to different internal (inside the enterprise) and external services. The proposed process is intended for the development of the enterprise portals. It is a generic and global methodology, i.e. it is not specified for a particular type of portals.

The main objective of this process is the integration of the existing application through the use of the web services.

It describes how to transform the existing applications into a set of services and to add new components developed inside the company and the external services. This process is iterative and it follows the RUP process i.e. the phases are carried out in a repetitive way until that the all requirements are accomplished. It is divided into six steps (fig.1): specification, analysis, re-use of the existing applications, design of the user interfaces, implementation and management.

### 3. 1 Global specification phase

The goal of this phase is to determine the orientation of the project. In other words, the general objective is to give the broad outline of the portal in order to determine the list of the internal and external systems that will be interact with the enterprise portal and to conclude which are the various users of this last.

Generally, the internal systems (sub systems) are the departmental Intranets of the enterprise. Each one can contain one or several applications, which use various data sources. The external systems are generally provided by the partner companies. These systems are accessible through an Extranet or Internet, as they can be public web sites, which provide useful services and functionalities for the enterprise (for example, a purse service).

It is also necessary to determine the needs of the enterprise with giving answers to several questions:
- What is the objective of the portal that the enterprise wants to develop?
- What are the future users of the portal?
- What are the services which the company wants to provide to the users?
- What are the partners that will participate in the project?

At the end of the specification phase, the designer can determine and evaluate the complexity of the project.

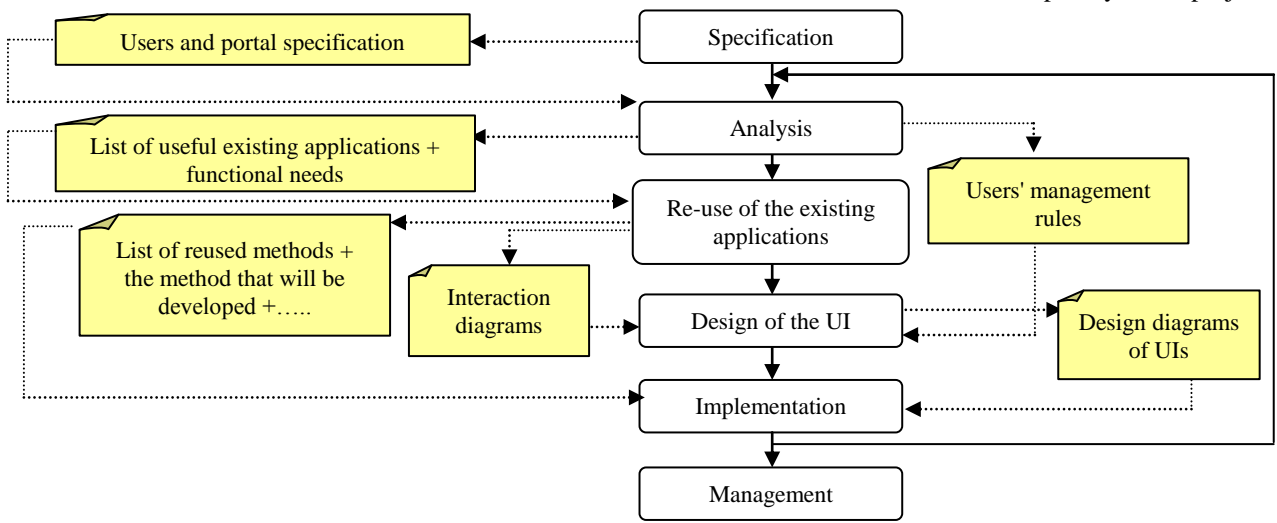

Figure 1. Different steps of the development process

## 3. 2 Analysis phase

After the determination of the type of portal and the various types of the future users in the previous step, this phase has two principal objectives: the definition of the functional needs and the management of the users.

### 3. 2.1 Definition of the functional needs

During this step, one can define the global contents of the portal. After the determination of the basic list of the different services, the dialogue with the users and the observation of the daily activities, enable us to enrich the list of the services. Several interesting information can be recovered for this step, for instance: the list of the internal applications those are useful for the project, technologies with which these applications have been developed, categories of users and management rules used in each internal system (or sub system).

### 3.2.2 Users management

After the determination of the portal contents, one can process very important aspects related to the operational requirements of portal, the access control and user authentication. All these functions are essential factors for the personalization and the creation of the user profiles. Moreover, in this stage we can specify the list of services, which are authorized for each type of user.

## 3. 3 Re-use of the Existing Applications

With starting from the basic list of the internal services that will be used, it is necessary to compare the functionalities provided by the existing applications; those are considered useful for the project and the awaited functionalities. This comparison enables us to specify the services to be added (software components, which realize certain missing functionalities). In order to determine the relation between the internal services that will be integrated in the portal and the existing applications, it is necessary to study the internal services (current needs) independently of what exists, and to compare the result obtained with the existing applications.

The objective of this phase is to create a conceptual model for each service. This model is represented with a set of objects that are characterized by a name, attributes and operations. Then a comparison of the operations related to the found objects with those provided by the existing applications is made. If these operations are automated in an existing application, then they will be re-used. For that, we call use cases, class and object diagrams of UML.

### 3.3.1 Obtaining of Use cases

The use cases are useful to express the interaction of the system with the different actors (user or other systems) and to determine the various classes of the conceptual model. Generally, they are obtained in an incremental way i.e. the designer adds at each time the use cases and actors to derive all the conditions and the actions which the system will have to execute.

### 3.3.2 Description of the Use Cases

Before designing the conceptual model, the description of the use cases by scenarios is very necessary. This last can be carried out in a textual way or in the form of interaction diagrams (sequence or collaboration diagrams).

### 3.3.3 Definition of the conceptual model

During the definition of the conceptual model, it is necessary for the designer to use the documents that have already been obtained during the development of the existing application. The comparison is easier if the old design is carried out with an oriented object method, which uses the UML notation. A second possibility of comparison is carried out between the conceptual model and the implementation of the application. So, the comparison between the design model of UML (primarily the conceptual model and interaction diagrams) and the existing operations is made.

Once the comparison is made, we can determine four possibilities:

1. Among the applications that we chose to integrate in the portal, there are those that we can re-use directly, i.e. the functionalities of the applications that already exist are exposed like a web service. This possibility is realizable, if the existing application is implemented in a language which is compatible with the web service technology like C#, vb.net, java, and so on.

2. Some applications are not directly reusable. These applications require a phase of transformation (migration) or wrapping. The translation can be carried out manually and/or automatically from the source code into another language (java for example). However, the wrapping operation consists to create a transition layer (Wrapper), which allows the interaction with the source code. A good example is the creation of a java wrapper that uses Java Native Interface (JNI). We can apply this approach to FORTRAN, C and C++ code.

3. If the method does not exist, it will be developed.

4. The fourth possibility is that the awaited web services will be composed from other ones. This case requires the development of a workflow engine. If the workflow system already exists, this situation requires a study for the integration of a service in this system.

### 3.4 Design of user interfaces

In the previous phase, we described the development process of the web services for provider side; in this step we explain how to realize the application of user side to consume the available web services (internal and external). Practically, this phase is decomposed into two principal levels:

The first level is used to enrich the previous phases. Thus, we design the dynamic part of the system independently of the platform, which will be used for the implementation. To complete this task we use the sequence diagrams, activity diagrams and the collaborative diagrams.

For the external services, generally, its presentations are provided by the service providers (in the form of portlet). In this case, these presentations will be integrated directly in the portal, if they are compatible with the portal framework. In other cases, we need to develop local presentations (to measure) for certain external services.

The second level of this phase makes the relation between the result of the first part and the software components used to implement the various presentations. In order to facilitate the task and to express the relationship with the implementation of the presentations, a programming pattern of the web applications should be used. The majority of current works propose the use of the MVC (Model, View, Controller) paradigm for designing the web applications, because it presents an important advantage which is the separation between the application model and the user interface.

### 3.5 Implementation Phase

The implementation phase is divided into two parts: the development of the internal services and the implementation of the customer application. The internal services are expressed from the existing applications (using the result of the reuse of existing applications phase).

### 3.6 General Management

This phase contains the management and the administration of the portal. This function is not limited by a precise time, but it must be continuous to answer to the enterprise evolutions.

## 4. A motivate example

In this section, we present a simplified example relating to an application that manages customer commands. Our objective is to re-use this application in the form of web services which will be integrated in an e-business portal.

### 4.1 Obtaining of Use cases

In figure 2, we present a use case diagram for a selling service. This diagram is very simplified because the actions carried out by the actor include under actions.

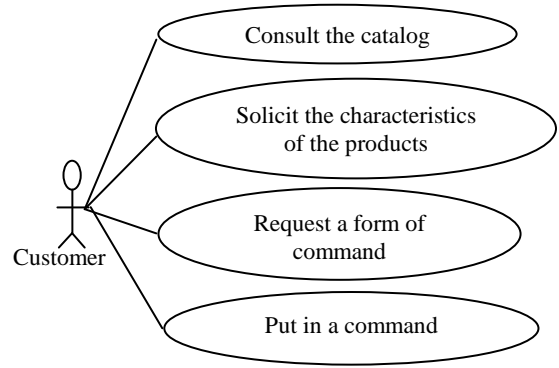

**Figure 2.** Example of a selling service (simplified view).

### 4.2 Description of the Use Cases

For example, we describe the use case "put in a command" in a textual way as follows:

*Pre conditions*
- The customer already consulted the available products catalogue.
- The customer already recovered the characteristics of the products.

*Scenario*
- The customer asks for a command form;
- The system sends the command form to the customer;
- The customer fills the command form;
- The customer orders the command;
- The system validates the provided information;
- The system saves the command;
- The system sends a confirmation to the customer;
- The customer reads the confirmation.

*Post conditions*
- It is obligatory that the customer receives the confirmation.

### 4.3 Definition of the conceptual model

The designer can extract the classes starting from the scenarios and of the interaction diagrams. He can be carried out a syntactic analysis of the various scenarios. In addition to the scenarios, other elements play a part to carry out the conceptual model, such as the management rules that are very useful to describe the associations between the classes and to define the cardinalities.

After the comparison between the conceptual model and the existing operations, we can decide which are the names and the parameters of the methods which will be normally implemented. Finally, we compare this whole of method with the operations, which already exist (existing applications).

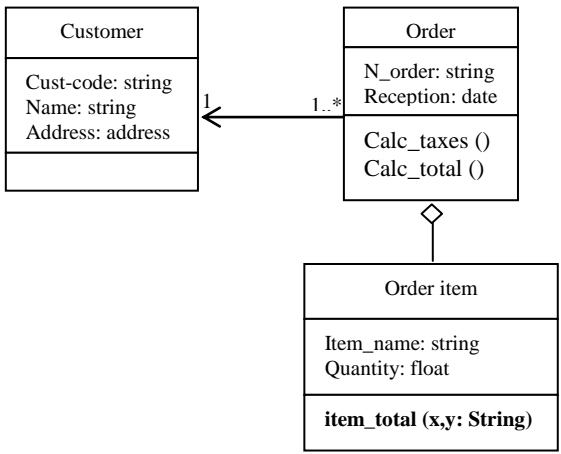

**Figure 3**. A simplified conceptual model of a selling service

### 4.4 Implementation of the web service

For example, the function **item_total(x,y:String)** (appeared in the conceptual model of figure 3) belongs originally to an existing application developed in VB6. We re-used this application in a way where we carried out an operation of migration (automatically) towards VB.NET, and then we exposed this function like a web service. For this, we have used the Visual Basic Upgrade Wizard of visual studio .NET.

The following VB.NET code shows the function **item_total (x,y:String)**:

```
Imports System.Web.Services
Imports System
Imports System.Data
Imports System.Data.OleDb
………………………………………………
<WebService(Namespace := "http://tempuri.org/")> _
Public Class Order_item
Inherits System.Web.Services.WebService
...............................................................
<WebMethod (Description:=" display the quantity requested and the total price of an item ",EnableSession:=False)>
Public Function item_total (ByVal X, Y As String) As DataSet
...............................................................
End Function.
```

### 4.5 Implementation of the presentation (portlet)

For the requestor role, there are three principal steps:
- Generation of a proxy class from the WSDL description of the service.
- Development of a portlet (presentation of the service) that uses the proxy class to invoke the methods exposed by the service.
- Integration of the portlet in a portal framework that we comply with the users' management. For example, we can use an open source like Apache Jetspeed Enterprise Information Portal [15] (fig 4).

The following java code shows the implementation of a Jetspeed Portlet which invoke the method **item_total (x,y:String)** exposed by the web service "selling_service.asmx".:

```
import org.apache.ecs.*;
import org.apache.ecs.html.*;
Import org.apache.jetspeed.portal.portlets.AbstractPortlet;
………………………………………… …………………
import Sproxy.*;
……………………………………………………………………
public class clientproxy {
……………………………………………………………………
Stub stube = createProxy();
 stube._setProperty(javax.xml.rpc.Stub.
ENDPOINT_ADDRESS_PROPERTY,
"http://domaine.com/selling_service.asmx");
Sproxy.selling_serviceSoap service =
Sproxy.selling_serviceSoap) stube;
……………………………………………………………………
System.out.println (service.item_total (NO,IN));
```

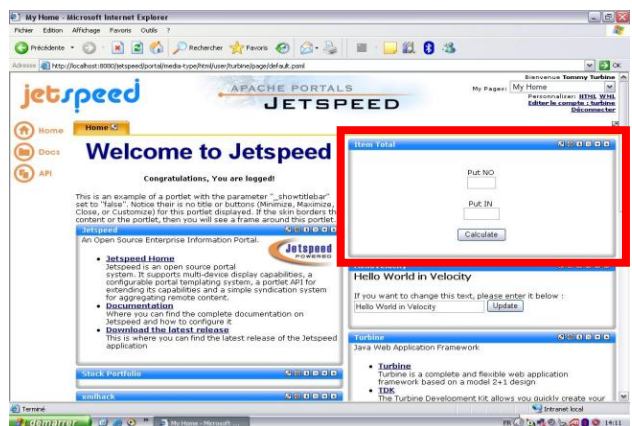

**Figure 4**. Item total Portlet .

## 5. Some related works

Web services and enterprise portals are two new Information technologies (IT) that gathered in several important recent researches. Pierce & al [11] present a view of computing portal architectures based around web services model, with a description of the initial investigations into the core services, interoperability issue, and security. A generic portal framework for integration of existing expertise and services of individual institutions is presented in [4]. This paper discusses the presentation-level application integration with workflow support. In particular, it explores the real and rich scenarios found in e-learning where education services are offered through the Internet by networked universities. Guillermo and Javier

[16] propose an architectural approach for an e-Services Hub to allow Small and Medium Enterprises share technological competences. Moreover, a prototype implementation of the software components of the e-Services Hub was implemented.

In our work, we discuss some aspects which are already studied in preceding works, like the separation between the interface and the service. Moreover, we present the methodological aspect for the development of the web service portals. In addition, we describe how to re-use the existing applications (including the legacy systems) to build new services. It's the more important feature that differentiates our approach from theirs.

In this paper, we describe a generic and global methodology for the development of the web services for the enterprise portals by re-using existing applications. In our context we consider that an enterprise portal is an integration project i.e. the portal is a hub for internal and external web services.

## 6. Conclusion

In this paper, we described a development process of an enterprise portal, which integrates and aggregates a set of web services. This process is based on two levels. The first level concerns the preparation and the construction of the global contents integrated by the portal. The second level describes the development of the portal. For the first part, we added the phase "Re-use of the existing applications" that makes the binding between the Web services (the present needs for the users) and the already existing applications. The objective of this phase is to quickly provide the elements of the decision for the services that will be extracted from the existing applications and those which will be developed.

The second part shows how to develop the portal that is a web application. For this reason, we reuse some aspects used in the RUP like the guiding of the process by the use cases. But there remain some aspects that are not studied in this work especially those that are related to the business processes of the enterprise. In this case, in more with SOAP, UDDI and WSDL we can use other standards dedicated to the realization of Workflow like BPEL4WS (Business Process Execution Language for Web Services).

In addition, we try defining the passage rules to carry out the comparison between the conceptual model and the existing applications.

Moreover, to develop the search engine of the portal, we can be benefited from semantic web and semantic web services technologies to find the functionalities of the existing services (especially external ones).

Finally, to call the external Web services, we can use the standard WSRP (Web Services for Remote Portlet) [14]. In this case, the various developed presentations (portlet) are published to be used directly by the external consumers. In the same way, the enterprise can integrate directly in its portal the presentations published by other remote portals.